\documentclass[structabstract]{aa}  
\usepackage{graphicx}
\usepackage{txfonts}
\usepackage{natbib}
\begin{document}
%

  \title{QSO Selection and Photometric Redshifts with  Neural Networks}

  \author{
 	Ch. Y\`eche\thanks{christophe.yeche@cea.fr}\inst{1}
        P. Petitjean\inst{2}, J. Rich\inst{1},
        E. Aubourg\inst{3}, N. Busca\inst{3}, J.-Ch. Hamilton\inst{3},
        J.-M. Le Goff\inst{1},  I. Paris\inst{2}, S. Peirani\inst{2}, Ch. Pichon\inst{2}, E. Rollinde\inst{2},      
         \and M. Vargas-Maga\~na\inst{3}
         }

  \institute{CEA, Centre de Saclay, IRFU,  F-91191 Gif-sur-Yvette, France,  
            \and    
            Universit\'e Paris 6 et CNRS, Institut d'Astrophysique de Paris, 98bis blvd. Arago, F-75014 Paris,  France,
             \and
             APC,  10 rue Alice Domon et L\'eonie Duquet, F-75205 Paris Cedex 13, France
            }



 \abstract
  {
  Baryonic Acoustic Oscillations (BAO) and their effects on the matter power spectrum
can be studied by using the Lyman-$\alpha$ absorption signature of the matter
density field along quasar (QSO) lines of sight.
A measurement sufficiently accurate to provide useful cosmological constraints
requires the observation of $\sim10^5$ quasars in the redshift range $2.2<z<3.5$
over $\sim8000{\rm deg^2}$.
Such a survey is planned by the Baryon Oscillation Spectroscopic Survey (BOSS)
project of the Sloan Digital Sky Survey (SDSS-III).
}
  {One of the challenges for this project is to build from five-band imaging data
a list of targets that contains the largest number of quasars in the required redshift 
range.
In practice, one needs a stellar rejection of more than two orders 
of magnitude with a selection efficiency for quasars better than 50\%
up to magnitudes as large as $g\sim 22$. 
Standard methods to identify quasars using colors work well for brighter quasars in the range
0.3~$<$~$z$~$<$~2.2 and $g<21$ but it is necessary to develop new methods for higher
redshifts and magnitudes. 
}
  {To obtain an appropriate target list and estimate quasar redshifts, 
we have developed an Artificial Neural Networks (NN) with a multilayer perceptron 
architecture. The input variables are photometric measurements, i.e. the object 
magnitudes and their errors in the five bands ($ugriz$) of the SDSS photometry. 
The NN developed for target selection  provides a continuous output variable
between 0 for non-quasar point-like objects to 1 for quasars.
A second  NN  estimates the QSO  redshift $z$ using the photometric information.}
  {For target selection, we achieve a non-quasar point-like object rejection of 
99.6\% and 98.5\% for a quasar efficiency of, respectively, 50\% and 85\%. The 
photometric redshift precision is of the order of 0.1 over the region relevant for BAO studies. 
These  statistical methods, developed in the context of the BOSS project, can easily be 
extended to any quasar selection and/or determination of their photometric redshift.  }
  {}


\authorrunning{Ch. Y\`eche et al.}
  \maketitle
%

\section{Introduction}

Since the first quasar was discovered \citep{schmidt63}, 
methods have been developed to
 differentiate these rare objects from other astronomical sources
 in the sky. In the standard methods, it is assumed that QSOs have
 point-like morphology. They are then separated from the much more
 numerous stars by their photometric colors. 
 The UVX selection, e.g. \citep{croom01}, 
 can be largely complete ($>$90\%) for QSOs with
 0.3~$<$~$z$~$<$~2.2 but this completeness drops at higher redshift.
 The selection purity was brought up to 97\% for $g<21$ using Kernel
 Density Estimation  techniques applied to SDSS colors \citep{kde04}
  and extended to the infrared by \citet{richards09} 
 implying that spectroscopy is not needed to
 confirm the corresponding statistical sample of quasars
at high galactic latitudes. This led to
 the definition of a one-million-QSO catalog \citep{kde09}
 down to $i=21.3$ from the photometry of 
SDSS Data Release 6 \citep{adelman08}.

Extending quasar selection methods to
higher redshifts and magnitudes presents several difficulties. 
For example, at fainter
 magnitudes, galaxies start to contaminate ``point-like'' photometric
catalogs both because of increasing photometric errors and because
of non-negligible contributions of AGN's in certain bands.
Nevertheless, such an extension is very desirable, not only to
study the AGN population
but also to use the quasars
to study the foreground absorbers. In particular
 studies of spatial correlations in the IGM from the Lyman-$\alpha$
 forest and/or metal absorption lines  are in need of higher target
 density at high redshift 
\citep{petitjean97,nusser99,pichon01,caucci08}.

More recently, it was realized that the Baryonic Acoustic Oscillations (BAO) 
could be detected in the Lyman-$\alpha$ forest. BAO 
in the pre-recombination Universe imprint features in the matter power spectrum
that have led to important constraints on the cosmological parameters.
So far, BAO effects have been seen using galaxies of redshift $z<0.4$
to sample the matter  density
\citep{eisenstein,cole,percival}.
The Baryon Oscillation Spectroscopic Survey (BOSS) \citep{boss} of the Sloan Digital 
Sky Survey (SDSS-III) \citep{sdss3} proposes to extend these studies using galaxies
of higher redshifts, $z<0.9$.
The BOSS project will also study BAO effects in the range $2.2<z<3.5$  using
Lyman-$\alpha$ absorption towards high 
redshift quasars (QSOs) to sample the matter density
as proposed by \citet{McDonald}.

\begin{figure*}[htb]
  \centering
  \includegraphics[width=14.0cm]{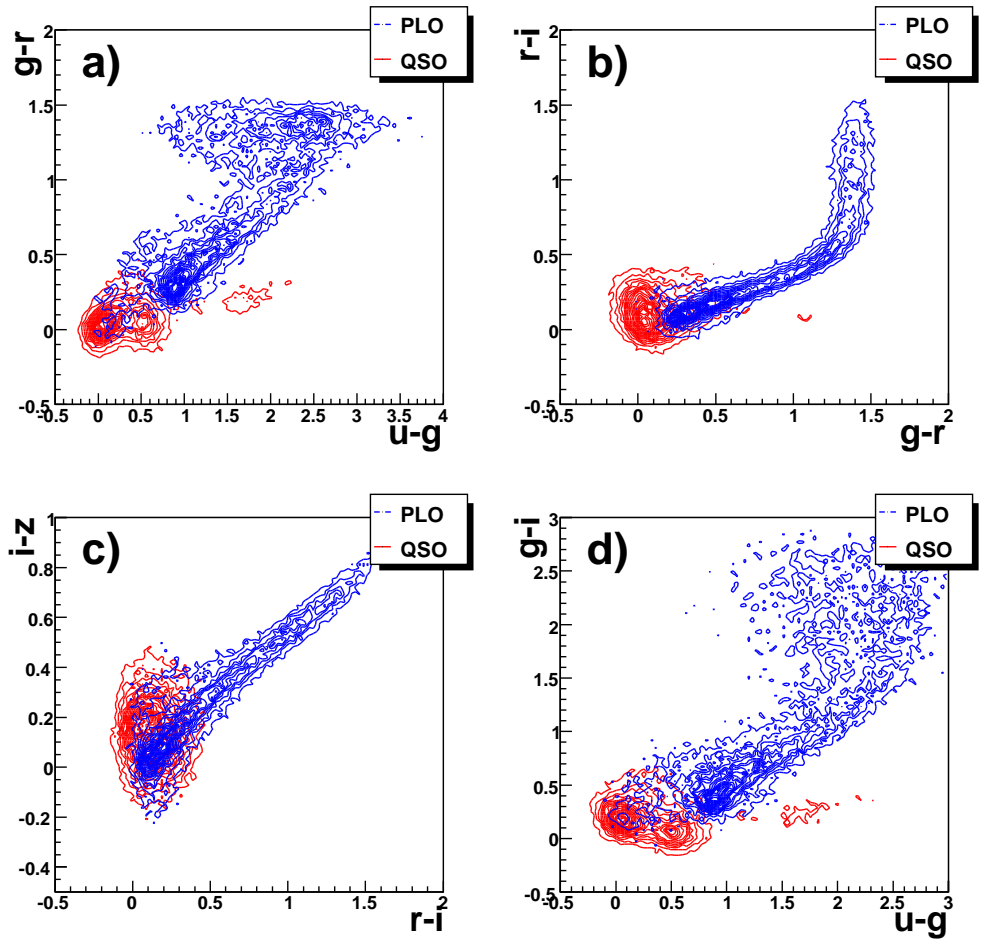}
     \caption{2D distributions of colors ($u-g$, $g-r$, $r-i$, $i-z$ and  $g-i$)  for 
     objects classified as PLO in  SDSS photometric catalog
 (blue lines for contours) and for  objects spectroscopically classified as QSO 
(red solid lines for contours). The PSF magnitudes ($ugriz$) have been corrected for
Galactic extinction according to the model of \citet{schlegel98}. 
}
        \label{fig:Colors}
  \end{figure*}

\begin{figure*}[htb]
  \centering
     \includegraphics[width=14.0cm]{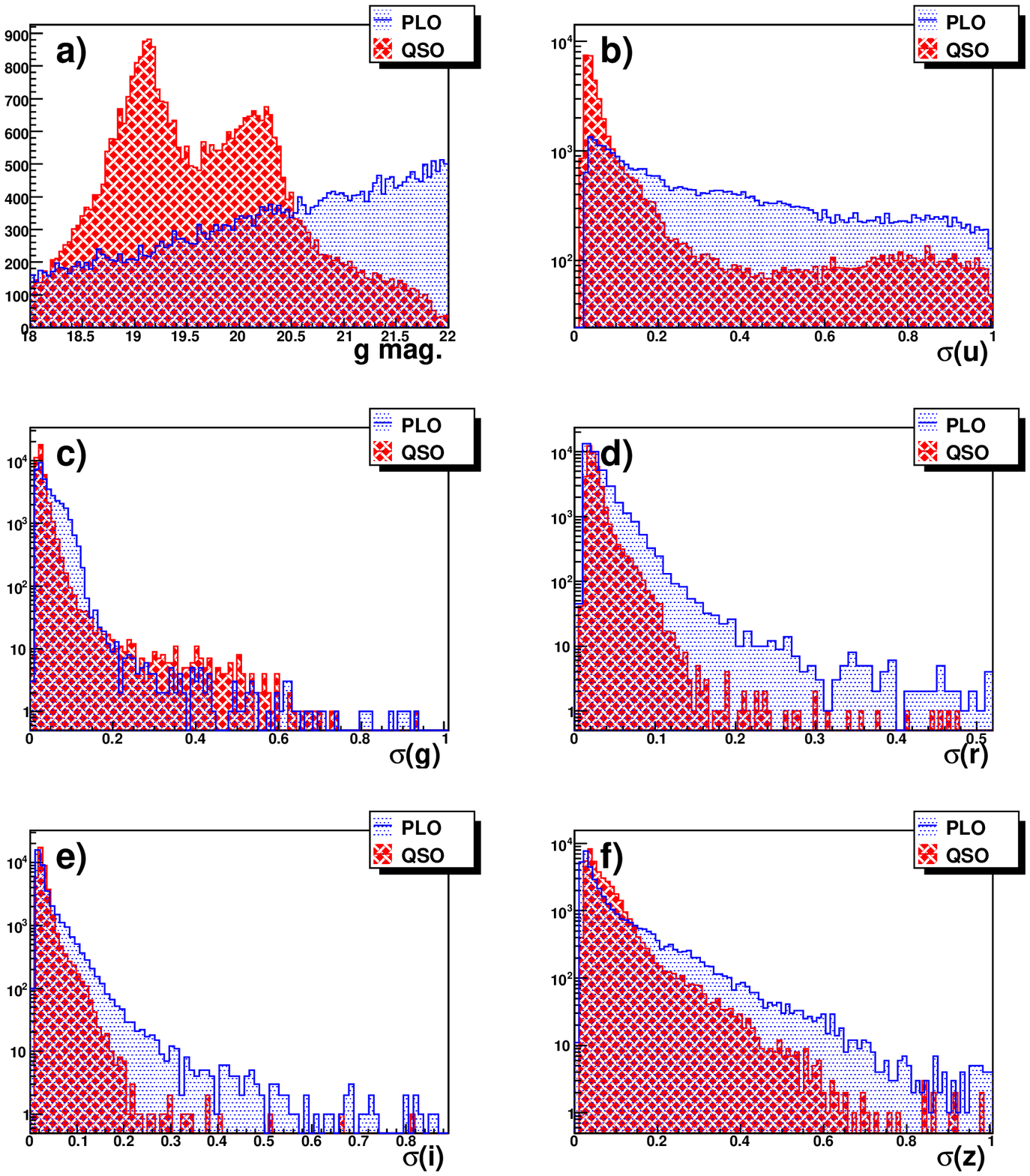}
     \caption{Distributions of the discriminating variables used as input in the NN for 
     objects classified as PLO in  SDSS photometric catalog
(blue dotted  histogram) and for  objects spectroscopically classified as QSO (red slashed
histogram):  {\bf a)} Distribution of the PSF $g$ magnitude, {\bf b), c), d) e) and f)} 
Distributions of, respectively, $\sigma(u)$, $\sigma(g)$, $\sigma(r)$, $\sigma(i)$ and 
$\sigma(z)$, the errors on the corresponding PSF magnitudes.
}
        \label{fig:compVar}
  \end{figure*}

The power spectrum has already been measured at $z\sim2.5$
via the 1-dimensional matter power spectrum derived from quasar spectra \citep{croft}.
The observation of BAO effects will require a full 3-dimensional sampling
of the matter density, requiring a much higher number of quasars
than previously available.
BOSS aims to study around 100,000 
QSOs over 8,000 square degrees.
The requirement that the Lyman-$\alpha$ absorption fall in the
range of the BOSS spectrograph requires that the quasars 
be in the redshift range $2.2<z<3.5$. 

The quasars to be targeted  must be chosen using only available photometric information,
mostly from the SDSS-I point-source catalog.
The target selection method must be able to reject the non-quasar point-like objects 
(PLOs; mainly stars) by more than two orders of magnitude with a selection 
efficiency of QSOs better than 50\%. The BOSS project needs a high 
density of $z>2.2$ fainter QSOs ($\sim 20$ QSOs per
 sq degree) and therefore requires the selection to be pushed up to
 $g\sim 22$. We developed a new method to select quasars using
 more information than the standard color selection methods.

The classification of objects is a task that is
generally performed by applying cuts on various distributions which 
distinguish signal objects from background objects.
This approach is not  optimal  because
all the information (the shapes of the variable distributions, the correlations between
the variables) is 
not exploited and this leads to a loss in classification efficiency.
Statistical methods based on multivariate analysis have been developed 
to tackle this kind of problem.
For historical reasons these methods have been focused on linear problems which are easily 
tractable. In order to deal with nonlinearities, Artificial Neural Networks (NN) have been 
shown to be a
powerful tool in the classification task (see for instance \citet{bishop}). 

By combining photometric measurements such as the magnitude values and their errors for the
five bands ($ugriz$) of SDSS photometry, a NN approach will allow us both to select the QSO
candidates and to predict their redshift. Similar methods such as Kernel Density Estimation (KDE) 
\citep{kde04,kde09} already exist to select QSOs.  Our approach based on NN is an extension 
of these methods because we  will use more information (errors and
absolute magnitude $g$ instead of only colors (difference between two magnitudes)).
Moreover, we propose to  treat in parallel the determination of the redshift  with
the same tool. This approach contrasts with the usual  methods to compute photometric redshift
which deal with $\chi^2$ minimization techniques  \citep{photoz,weinstein04}.

\section {QSO and Background Samples}

The quasar candidates should be selected among a photometric catalog of 
objects including real quasars and what we will call background 
objects.
Here, both for the background and QSO samples, the photometric information comes from 
the SDSS-DR7 imaging database of point-like objects \citep{dr7}, PLOs. 
We apply the same quality cuts on
the photometry for the two samples and select objects with $g$ magnitude in the
range $18 \leq g  \leq 22$. Note that in the following,
magnitudes will be point spread function (PSF) magnitudes 
\citep{lupton99} in the SDSS pseudo-AB magnitude system \citep{oke83}.   

\subsection{Background Sample}
\label{sec:qsosample}
For the background sample, we would ideally use an unbiased sample of 
spectroscopically confirmed SDSS point-like objects \emph{that are not QSOs}.
Unfortunately, we have no unbiased sample of such objects because spectroscopic
targets were chosen in SDSS-I to favor particular types of objects.
Fortunately, the number of QSOs among PLOs is sufficiently small that using
all PLOs as background does not affect the NN's ability 
to identify QSOs. 
We have verified that this strategy works by using the synthetic PLO 
catalog of \citet{fan}. We degraded the star sample by adding a few percent of QSOs in it. 
then, we retrained the NN and we  compared the NN trained with a pure star sample. 
We did not observe any significant worsening of the NN performances.

The background sample used in the following was drawn from the SDSS PLO sample.
We used objects with galactic latitude $b$ around $45^\circ$ to
average the effect of Galactic extinction. In the future, we may consider the possibility 
of having a different NN for each stripe of constant galactic latitude. The final sample 
contains ~30,000 PLOs:
half of them constituting the ``training" sample, the other half  the ``control"
sample, as explained in the next section.

\subsection{QSO Sample}
For the QSO training sample, 
we use a list of 122,818 
spectroscopically-confirmed quasars obtained from the 2QZ quasar catalog 
\citep{croom04}, the SDSS-2dF LRG and QSO Survey (2SLAQ)  \citep{croom09},
and the SDSS-DR7 spectroscopic database \citep{dr7}. These quasars have redshifts 
in the range $  0.05 \leq z  \leq 5.0 $ and $g$ magnitudes  in the  range
$18 \leq g  \leq 22$ (galactic extinction corrected). Since quasars will
be observed over a limited blue wavelength range (down to about 3700~\AA),
we will target only quasars with $z>2.2$.
Therefore, the sample of known quasars includes 
33,918 QSOs with $z\geq 1.8$: half of them constituting the effective
``training" sample, the other half  the 
``control" sample. For the determination of the photometric redshift, we use a wider sample of
95,266 QSOs  with $z\geq 1$.

In order to compare together QSOs with background objects from different regions of 
the sky, the QSO magnitudes have been corrected
for Galactic extinction with  the model of \citet{schlegel98}.

\subsection{Discriminating variables}

The photometric information is extracted from the SDSS-DR7 imaging database 
\citep{dr7}. The 10 elementary variables are the PSF magnitudes
for the 5 SDSS bands ($ugriz$) and their errors. As explained in \citet{kde09},
the most powerful  variables are the four usual colors
($u-g$,$g-r$,$r-i$,$i-z$) which combine the PSF magnitudes. Fig.~\ref{fig:Colors} shows the
2D color-color distributions  for the QSO and PLO samples. 

These plots give the impression that it is easy to disentangle the two classes of objects
but one needs to keep in mind that the final goal is to obtain a 50\% efficiency for
QSOs with a non-quasar PLO efficiency of the order of $\sim10^{-3}$. Therefore to improve the
NN performances, we added the absolute magnitude $g$ and the five magnitude errors. Their
distributions for the two classes are given on Fig.~\ref{fig:compVar}. An improvement
can be expected from the additional variables and also from the correlations between the
variables. Indeed, for example, it is expected that errors be larger for compact galaxies
compared to intrinsic point-like objects.

Note that the $g$ distribution for the QSOs is likely to be biased by the spectroscopic
selection. This issue will be addressed in the future with the first
observations of BOSS. Indeed the photometric selection of QSOs for these first
observations is based on loose selection criteria and it should provide a ``less biased" 
catalog of spectroscopically confirmed quasars, close to completeness up to $g=22$. 

\section{Neural Network Approach}

\begin{figure}[htb]
  \centering
 \includegraphics[width=7.cm]{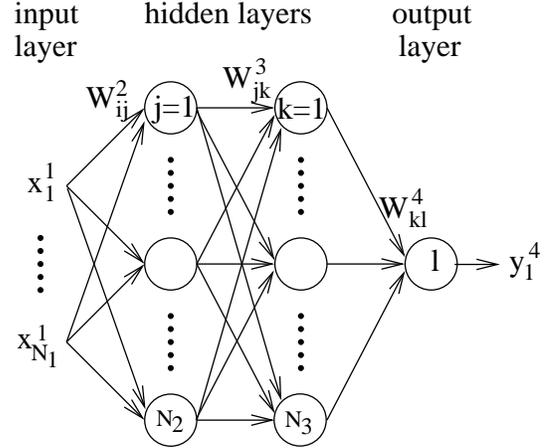}
     \caption{ Schematic representation of the Neurone Network used here with 
$N_1$  input variables, two hidden layers and one output neuron.}
        \label{fig:ArchitectureNN}
  \end{figure}

The basic building block of the NN architecture \footnote{
For this study, both for target selection and redshift determination, we use a C++ package,
TMultiLayerPerceptron developed in the ROOT environment  \citep{root}.}
is a processing element called a neuron.
The NN architecture used in this study is 
illustrated in Fig.~\ref{fig:ArchitectureNN} where
each neuron is placed on one of four ``layers'', 
with $N_l$ neurons in layer $l,\,l=1,2,3,4$.
The output of each neuron on the first (input) layer is 
one of the $N_1$ variables defining an object,
e.g. magnitudes, colors and uncertainties.
The inputs of neurons on subsequent layers ($l=2,3,4$) are the $N_{l-1}$ outputs of the
previous layer, i.e. the $x^{l-1}_j ,\,j=1,..,N_{l-1}$.
The inputs of any neuron are first linearly
combined according to ``weights'', $w^l_{ij}$ and ``offsets''
$\theta^l_j$ 
\begin{equation}
y^l_j=\sum_{i=1}^{N_l} w^l_{ij}\, x^{l-1}_i +  \theta^l_j\,\,
\hspace*{5mm}l\,\geq\,2 \;.
\end{equation}
The output of neuron $j$ on layer $l$ is then defined by the 
non-linear function
\begin{equation}
x^{l}_j 
=
\frac{1}{1+ \exp\left(-y^l_j\right)}\,\,
\hspace*{5mm} 2\leq \,l\,\leq 3 \;.
\label{eq:activation}
\end{equation}

The fourth layer has only one neuron giving an output  $y_{NN}\equiv y^4_1$,
reflecting the likelihood
that the object defined by the $N_1$ input variables is a QSO.              

 \begin{figure*}[htb]
  \centering
  \includegraphics[width=\textwidth]{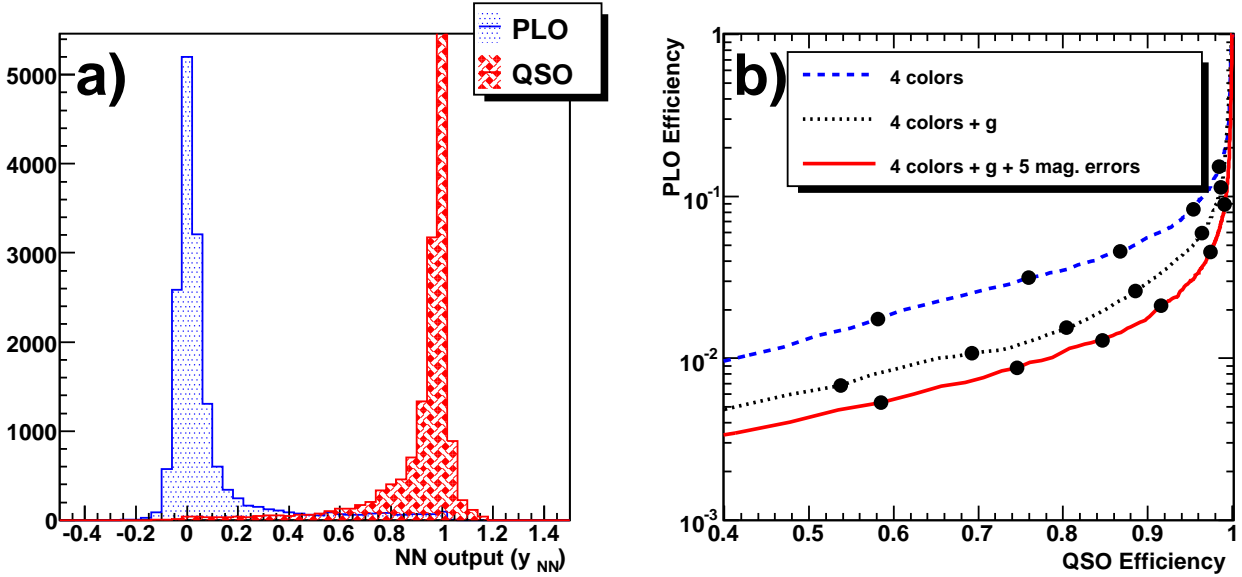}
     \caption{{\bf a)} NN output for  objects classified as PLO in the SDSS photometric catalog, 
i.e. background objects, (blue dotted histogram) and for objects 
spectroscopically classified as QSO (red slashed histogram) in the control samples,
using 10 discriminating variables: 4 colors, $g$ magnitude and errors on the 
five ($u,g,r,i$ and $z$) magnitudes.
{\bf b)} PLO efficiency 
as a function of the 
QSO efficiency 
for three NN configurations. 
Blue dashed line: 4 colors ($u-g,g-r,r-i,i-z$). Black dotted line: 4 colors + $g$ magnitude. Red 
solid line: 4 colors + $g$ magnitude + errors on the five  ($u,g,r,i$ and $z$) 
magnitudes. The curves are obtained by varying the cut value, $y^{min}_{NN}$
for the two distributions of Fig.~\ref{fig:NNOutputEffi}-a. Efficiency is defined 
as the ratio of the number of objects with a NN output greater than $y^{min}_{NN}$ over 
the number of objects in the sample. The dots correspond, from left to right, 
to $y^{min}_{NN}$ equal to, respectively, 0.2, 0.5, 0.8, 0.9, 0.95 and 0.98. }
        \label{fig:NNOutputEffi}
  \end{figure*}

Certain aspects of the NN procedure, especially the number of layers and the 
number of nodes per layer, are somewhat arbitrary and are chosen by experience
and for simplicity.
On the other hand, the weights and offsets must be optimized so that
the NN output, $y_{NN}$,  correctly reflects the probability
that an input object is a QSO.        
The NN must therefore be ``trained''
with a set of objects that are known
to be QSOs or not QSOs (background objects).
More precisely, the weights and offsets are determined by minimizing the ``error'' function
\begin{equation}
E= \frac{1}{2n}\sum_{p=1}^{n}(y_{NN}(p)-y(p))^2\,\, ,
\label{eq:error}
\end{equation}
where the sum is over $n$ objects, $p$, and where $y(p)$ is a discrete value
defined as  $y(p)=1$ (resp. $y(p)=0$) if the
object $p$ is a QSO (resp. is not a QSO). In the case of the NN developed 
to estimate a photometric redshift, the targeted value $y(p)$ is a continuous value 
equal to the true spectrometric redshift, $z_{spectro}$. Note that in the NN
architecture used for this study, the activation function, defined in Eq.~\ref{eq:activation}, 
is not applied to the last neuron, allowing the output variable to vary in a range wider
than $[0;1]$.


In this kind of classification analysis, the major risk is the ``over-training" of the NN. It
occurs when the NN has too
many  parameters ($w_{ij}$ and $\theta_j$) determined by too few training objects. 
Over-training leads to an apparent increase in the classification efficiency
because the NN  learns by heart the objects in the training sample. 
To prevent such a behavior,
the QSO and background samples are split into two independent sub-samples, called ``training"
and ``control" samples. 
The determination of the NN parameters ($w_{ij}$ and $\theta_j$)
is obtained by minimizing the error $E$, computed over the QSO and background training samples. 
The minimization is suspended as soon as the error for the control samples stops decreasing 
even if the error is still decreasing for the training samples. We have 
followed this procedure both for the target selection and the determination of the 
photometric redshift.

The result of the NN training procedure 
is shown in Fig.~\ref{fig:NNOutputEffi}-a. The histograms 
of $y_{NN}$ for the control QSO and background samples are overplotted.  
Most objects have either $y_{NN}\sim 1$ (corresponding to QSOs) or 
$y_{NN}\sim 0$ (corresponding to background objects).
QSO target selection is achieved by defining a threshold value  $y_{NN}^{min}$ to
be chosen between $y_{NN}= 1$ and $y_{NN}\sim 0$.
The optimal value of the threshold is obtained by balancing the number of accepted QSOs 
against the number of accepted background objects.
A plot of the QSO efficiency vs. the background efficiency is shown in  
Fig.~\ref{fig:NNOutputEffi}-b.

\section {Photometric Selection of Quasar}
For illustration, we have considered three NN configurations that differ 
by the number of discriminating variables.
The first one uses only the four standard colors ($u-g$,$g-r$,$r-i$,$i-z$).
In the second configuration, we add the absolute magnitude $g$ and finally in 
the third configuration, the errors on the five PSF magnitudes are also taken into account.  
For each configuration, we have 
optimized the number of neurons in the hidden layers and the number of iterations in 
the minimization to get the best  ``PLO efficiency--QSO efficiency" curve. The three
curves are superimposed on Fig.~\ref{fig:NNOutputEffi}-b. Adding information, 
i.e discriminating variables, clearly improves the classification performances. 
For instance, for a QSO efficiency of 50\%, the PLO rejection fraction increases 
from 98.8\%, to 99.4\% and to 99.6\% when the number of variables increases  
respectively from 4 to 5 and to 10. In the region of QSO efficiency in which we 
want to work, between 50\% and 80\%, the PLO background is reduced by a factor 3 
by adding 6 variables to the four usual colors.
The small improvement found by using photometric errors may be due to a small 
contamination of the PLO catalog by compact galaxies.

It is therefore apparent that the 10-variable NN should be used
for the purpose of selecting quasars in any photometric catalog. 
In that case,
the PLO rejection factors are respectively, 99.6\%,  99.2\% and 98.5\% for QSO 
efficiencies of 50\%, 70\% and 85\%.

 \begin{figure*}[htb]
  \centering
  \includegraphics[width=15cm]{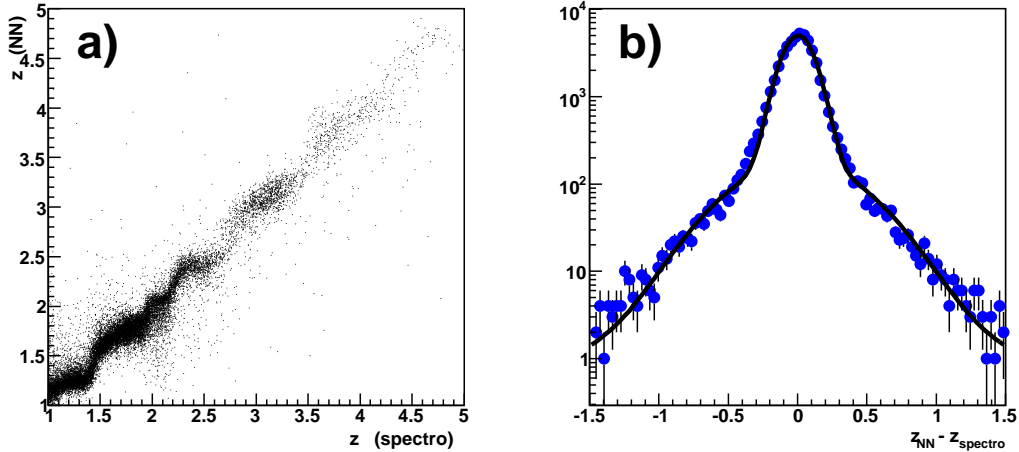}
     \caption{  {\bf a)} Photometric redshift determined with the NN ($z_{NN}$) as a function 
of the  redshift measured from spectroscopy ($z_{spectro}$).  {\bf b)} The $z_{NN}-z_{spectro}$ 
distribution is fitted with  three gaussians contributing $93.4\%$, $6.4\%$ and $0.2\%$ 
of the histogram
and of width, respectively, $\sigma$~=~0.1, 0.4 and 1.0.
The RMS of the $z_{NN}-z_{spectro}$ distribution is 0.18 and its mean is 0.00.}
        \label{fig:PhotozNN}
  \end{figure*}

According to the \citet{McDonald} computation based on the  
\citet{jiang} survey of faint QSOs, we expect $\sim$20 QSOs per deg$^2$, with  $g<22$ and
$2.2 \lesssim z \lesssim 3.5$. For a galactic latitude $b\sim45^\circ$, the number 
of objects selected in the SDSS-DR7 imaging database is $\sim$4000. Thus, with a QSO 
efficiency of 70\% and a PLO efficiency\footnote{Note that by its definition in Sec.\ref{sec:qsosample}, the PLO sample contains QSOs.} of 0.8\%, we will select 32 objects per deg$^2$ 
including $\sim$14  ``true" QSOs. These numbers corresponds roughly  to what is 
required for BOSS project. 

\section {Photometric Redshift of Quasar}

For the BOSS project, only quasars with a redshift in the range 
$2.2 \lesssim z \lesssim 3.5$ are useful. 
In the definition of the training sample, we have already applied a cut on 
the redshift, $z\geq 1.8$,  to reinforce the
selection of  high-$z$ QSOs. But it is useful to add an additional constrain and select 
only QSOs with $u-g>0.4$. This a-posteriori 
color cut helps to remove QSOs in the region $0.8 \lesssim z \lesssim 2.2$. However,
we propose a more elegant method which consists of estimating the redshift
of the QSO from the photometric information with another NN. 

For the determination of the photometric redshift we use the same 10 variables as those 
in the NN for target selection.  The difference is that in the definition of 
the error $E$, in Eq.~\ref{eq:error},
the targeted value $y(p)$ is a continuous value equal to the true spectrometric redshift, 
 $z_{spectro}$. Except for this difference, the NN architecture is the same as for target 
selection with two hidden layers with the same number of hidden neurons.
The minimization is computed with a single ``training" sample of spectroscopically-confirmed 
QSOs and it is suspended as soon as the error $E$ for the QSO ``control" sample   stops decreasing.

Fig.~\ref{fig:PhotozNN}-a shows the photometric redshift $z_{NN}$, determined with the NN versus
the spectroscopic redshift of the spectroscopically-confirmed QSOs. Most of the objects are
distributed along the diagonal demonstrating  the good agreement between the two measurements.
This can be quantified by plotting the difference $z_{NN}-z_{spectro}$
(Fig.~\ref{fig:PhotozNN}-b). The fit of this
 distribution with three gaussians gives   $93.4\%$ and
  $6.4\%$ of the objects respectively in  core and  wide Gaussians. The fraction
 of outliers, determined with the third Gaussian  is only $0.2\%$.

The corresponding distribution can be fitted 
with three Gaussian functions comprizing, respectively, $93.4\%$, $6.4\%$ and 0.2\%  of 
the distribution and of width, $\sigma$~=~0.1, 0.4 and 1.

\begin{figure}[htb]
  \centering
  \includegraphics[width=10cm]{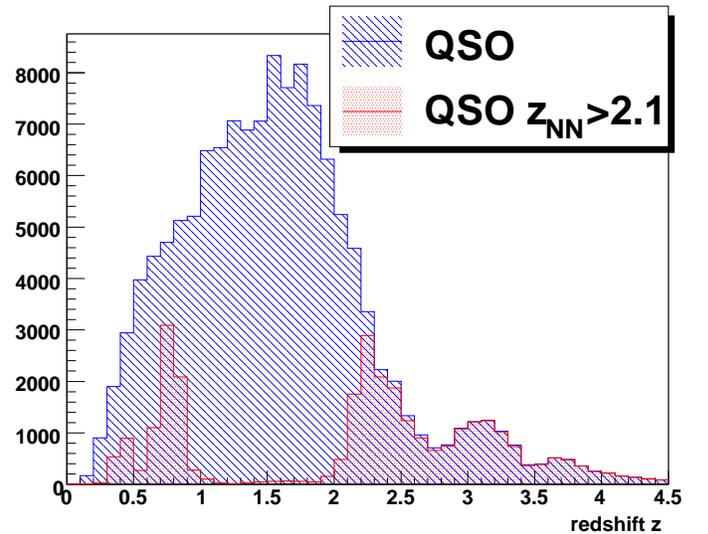}
     \caption{Spectrometric redshift distribution in the QSO sample (blue slashed histogram). 
The distribution for the QSO passing the cut $z_{NN}>2.1$ is overplotted (red dotted histogram). 
After this cut, 90.0\% of the QSOs with $z<2.2$ are removed and only 5.3\% of the QSOs in the $2.2< z <3.5$
region are lost. }
        \label{fig:QSORedshiftNN}
  \end{figure}

Therefore, as shown on Fig.~\ref{fig:QSORedshiftNN}, by applying a conservative cut on the
photometric redshift, $z_{NN}>2.1$, we can remove  90.0\% of the QSOs with $z<2.2$.
The fraction of lost QSOs with a redshift in the relevant region, $2.2< z <3.5$, stays
at a reasonable level of 5.3\%.  

\section{Conclusions}

In this paper we have presented a new promising approach to select quasars from
photometric catalogs and to estimate their redshift. 
We use an Neurone Network with a multilayer perceptron architecture. The input 
variables are photometric measurements, i.e. the magnitudes and their errors 
for the five bands ($ugriz$) of the SDSS photometry. 

For the target selection, we achieve a PLO rejection factor of 99.6\% and 98.5\% for, respectively,
a quasar efficiency of 50\% and 85\%.  The rms of the difference between the photometric redshift 
and the spectroscopic redshift is of the
order of 0.15 over the region relevant for BAO studies. These new statistical methods 
developed in the context of the BOSS project can easily be extended to any other analysis
requiring  QSO selection and/or  determination of their photometric redshift. 

\begin{acknowledgements}
     We thank N. P. Ross and D. H. Weinberg for triggering our interest 
     in QSO target selection in the context of the BOSS project and for many
      interesting discussions. The authors are also grateful to G. T. Richards, 
      A. D. Myers and E. Sheldon for important discussions and  
	for providing the QSO catalog  developed for the target selection in BOSS 
	and used in this paper. We like also to thank Fan X. who has provided us some synthetic
	catalogs of PLOs.
\end{acknowledgements}


\begin{thebibliography}{}

\bibitem[Adelman-McCarthy et al.(2008)]{adelman08} Adelman-McCarthy, J. K., et al., 2008, \apjs, 175, 297 
\bibitem[Abazajian(2009)]{dr7} Abazajian, K. N.  et al.,  (the Seventh Data Release of the SDSS),  2009, \apjs, 182, 543 
\bibitem[Bishop(1995)]{bishop} Bishop,  C. M., ``Neural Networks for pattern recognition,'' 1995, Oxford University Press  
\bibitem[Brun et al.(1995)]{root} Brun, R. et al., (the ROOT Team)http://root.cern.ch
\bibitem[Caucci et al.(2008)]{caucci08} Caucci, S. et al., 2008, \mnras, 386, 211
\bibitem[Cole et al.(2005)]{cole}  Cole, S.  et al., (the 2dFGRS Team),  \mnras, 362, 505 
\bibitem[Croft et al.(1999)]{croft} Croft, R. A. C. et al., 1999, \apj, 520, 1
\bibitem[Croom et al.(2001)]{croom01} Croom, S. M. et al.,  2001, \mnras, 322, 29
\bibitem[Croom et al.(2004)]{croom04} Croom, S. M. et al.,  2004, \mnras, 349, 1397
\bibitem[Croom et al.(2009)]{croom09} Croom, S. M. et al.,  2009, \mnras, 392, 19
\bibitem[Eisenstein et al.(2005)]{eisenstein} Eisenstein, D. J.  et al., (the SDSS Collaboration), 2005  \apj,  633, 560
\bibitem[Fan(1999)]{fan}  Fan, X. 1999, \aj, 117, 2528 
\bibitem[Jiang et al.(2006)]{jiang}  Jiang, X. et al., 2006, \aj, 131, 2788 
\bibitem[Lupton et al.(1999)]{lupton99} Lupton, R. H., Gunn J. E., Szalay A. S., 1999, \aj, 118, 1406
\bibitem[McDonald \& Eisenstein(2007)]{McDonald} McDonald, P. and Eisenstein, D. J., 2007, \prd, 76, 063009
\bibitem[Nusser \&Haehnelt, 1999]{nusser99}Nusser, A. \& Haehnelt, M., 1999, \mnras, 303, 179
\bibitem[Oke \& Gunn(1983)]{oke83} Oke, J. B., Gunn, J. E., 1983, \apj, 266, 713
\bibitem[Percival et al.(2009)]{percival} Percival, W. J. at al., 2009, submitted to \mnras, arXiv:0907.1660
\bibitem[Petitjean (1997)]{petitjean97} Petitjean, P., ``The Early Universe  with the
 VLT", edited by Jacqueline Bergeron 1997,  Berlin, Springer,  266
\bibitem[Pichon et al.(2001)] {pichon01}Pichon, C. et al., 2001, \mnras, 326, 597
\bibitem[Richards et al.(2001)]{photoz} Richards, G. T. et al., 2009, \aj, 122, 1151
\bibitem[Richards et al.(2004)]{kde04} Richards, G. T. et al., 2004, \apjs, 155, 257  
\bibitem[Richards et al.(2009a)]{richards09} Richards, G. T. et al., \aj, 137, 3884  
\bibitem[Richards et al.(2009b)]{kde09} Richards, G. T. et al., 2009, \apjs, 180, 67  
\bibitem[Schlegel et al.(1998)]{schlegel98} Schlegel, D. J., Finkbeiner, D. P., Davis, M., 1998, \apj, 500, 525
\bibitem[Schlegel, White \& Eisenstein(2009)]{boss} Schlegel, D., White, M., and Eisenstein, D.,  2009, arXiv:0902.4680
\bibitem[Schmidt(1963)]{schmidt63} Schmidt, J. A. 1963,  \nat, 197, 1040
\bibitem[SDSS-III Coll.(2008)]{sdss3} SDSS-III Collaboration, http://www.sdss3.org/collaboration/description.pdf
\bibitem[Shanno(1970)]{shanno} Shanno, D. F.,  1970, Math. Comp.,  24, 647
\bibitem[Weinstein et al.(2004)]{weinstein04} Weinstein, M. A. et al., 2004, \apjs, 155, 243 

\end{thebibliography}
\end{document}